\documentclass[apjl]{emulateapj}

\usepackage{amsmath}
\usepackage{graphicx}
\usepackage{epsfig,psfrag,epic,eepic}
\usepackage{textcomp}
\usepackage{mathptmx}

\slugcomment{Accepted for publication in ApJ Letters}

\shorttitle{Imaging Discovery of the Debris Disk Around HIP 79977}
\shortauthors{Thalmann et al.}

\newcommand{\pannekoek}{1}
\newcommand{\princeton}{2}
\newcommand{\arizona}{3}
\newcommand{\oklahoma}{4}
\newcommand{\charleston}{5}
\newcommand{\goddard}{6}
\newcommand{\toronto}{7}
\newcommand{\madrid}{8}
\newcommand{\subaru}{9}

\newcommand{\fizeau}{10}
\newcommand{\mpia}{11}
\newcommand{\lmu}{12}
\newcommand{\naoj}{13}
\newcommand{\ifahawaii}{14}
\newcommand{\tokyo}{15}
\newcommand{\sokendai}{16}
\newcommand{\jpl}{17}
\newcommand{\sinica}{18}
\newcommand{\physmath}{19}
\newcommand{\sapporo}{20}
\newcommand{\sendai}{21}

\newcommand{\noteone}{\ensuremath{^\mathrm{(1)}}}
\newcommand{\notetwo}{\ensuremath{^\mathrm{(2)}}}

\DeclareFontFamily{U}{euc}{}
\DeclareFontShape{U}{euc}{m}{n}{<-6>eurm5<6-8>eurm7<8->eurm10}{}%
\DeclareSymbolFont{AMSc}{U}{euc}{m}{n} 
\DeclareMathSymbol{\umu}{\mathord}{AMSc}{"16} 

\begin{document}

\title{Imaging Discovery of the Debris Disk Around HIP~79977\altaffilmark{$\star$}}

\author{C. Thalmann\altaffilmark{\pannekoek},   
	M. Janson\altaffilmark{\princeton},    
	E. Buenzli\altaffilmark{\arizona},    
	T.~D. Brandt\altaffilmark{\princeton},   
	J.~P. Wisniewski\altaffilmark{\oklahoma}, 
	C. Dominik\altaffilmark{\pannekoek},    
	J. Carson\altaffilmark{\charleston},    
	M.~W. McElwain\altaffilmark{\goddard},   
	T. Currie\altaffilmark{\toronto},
	G.~R. Knapp\altaffilmark{\princeton},   
	A. Moro-Mart\'in\altaffilmark{\madrid},    
	T. Usuda\altaffilmark{\subaru},   
	L. Abe\altaffilmark{\fizeau},   
	W. Brandner\altaffilmark{\mpia},   
	S. Egner\altaffilmark{\subaru},   
	M. Feldt\altaffilmark{\mpia},   
	T. Golota\altaffilmark{\subaru},   
	M. Goto\altaffilmark{\lmu},    
	O. Guyon\altaffilmark{\subaru},   
	J. Hashimoto\altaffilmark{\naoj},   
	Y. Hayano\altaffilmark{\subaru},   
	M. Hayashi\altaffilmark{\subaru},    
	S. Hayashi\altaffilmark{\subaru},    
	T. Henning\altaffilmark{\mpia},     
	K.~W. Hodapp\altaffilmark{\ifahawaii},   
	M. Ishii\altaffilmark{\subaru},   
	M. Iye\altaffilmark{\naoj},      
	R. Kandori\altaffilmark{\naoj},    
	T. Kudo\altaffilmark{\naoj},     
	N. Kusakabe\altaffilmark{\naoj},   
	M. Kuzuhara\altaffilmark{\naoj,\tokyo},    
	J. Kwon\altaffilmark{\sokendai,\naoj},   
	T. Matsuo\altaffilmark{\naoj},    
	S. Mayama\altaffilmark{\sokendai},   
	S. Miyama\altaffilmark{\naoj},    
	J.-I. Morino\altaffilmark{\naoj},   
	T. Nishimura\altaffilmark{\subaru},   
	T.-S. Pyo\altaffilmark{\subaru},   
	E. Serabyn\altaffilmark{\jpl},    
	H. Suto\altaffilmark{\naoj},    
	R. Suzuki\altaffilmark{\naoj},    
	M. Takami\altaffilmark{\sinica},   
	N. Takato\altaffilmark{\subaru},   
	H. Terada\altaffilmark{\subaru},   
	D. Tomono\altaffilmark{\subaru},   
	E.~L. Turner\altaffilmark{\princeton,\physmath},   
	M. Watanabe\altaffilmark{\sapporo},   
	T. Yamada\altaffilmark{\sendai},   
	H. Takami\altaffilmark{\subaru},   
	M. Tamura\altaffilmark{\naoj}   
}

\altaffiltext{$\star$}{Based on data collected at Subaru Telescope, which
	is operated by the National Astronomical Observatory of Japan.}

\altaffiltext{\pannekoek}{Astronomical Institute ``Anton Pannekoek'', 
	University of Amsterdam, Amsterdam, The Netherlands; \texttt{thalmann@uva.nl}.}
\altaffiltext{\princeton}{Department of Astrophysical Sciences, Princeton University, USA.}
\altaffiltext{\arizona}{Department of Astronomy and Steward Observatory, University of Arizona, Tucson AZ, USA.}
\altaffiltext{\oklahoma}{H.L.\ Dodge Dept.\ of Physics \& Astronomy, Univ.\ of Oklahoma, USA.}
\altaffiltext{\charleston}{College of Charleston, Charleston, South Carolina, USA.}
\altaffiltext{\goddard}{NASA Goddard Space Flight Center, Greenbelt MD, USA.}
\altaffiltext{\toronto}{University of Toronto, Toronto, Canada.}
\altaffiltext{\madrid}{Department of Astrophysics, CAB-CSIC/INTA, Madrid, Spain.}
\altaffiltext{\subaru}{Subaru Telescope, Hilo, Hawai`i, USA.}
\altaffiltext{\fizeau}{Laboratoire Hippolyte Fizeau, Nice, France.}
\altaffiltext{\mpia}{Max Planck Institute for Astronomy, Heidelberg, Germany.}
\altaffiltext{\lmu}{Ludwig-Maximilians-Universit\"at, Munich, Germany.}
\altaffiltext{\naoj}{National Astronomical Observatory of Japan, Tokyo, Japan}
\altaffiltext{\ifahawaii}{Institute for Astronomy, University of Hawai`i, Hilo, Hawai`i, USA.}
\altaffiltext{\tokyo}{University of Tokyo, Tokyo, Japan.}
\altaffiltext{\sokendai}{Graduate Univ.\ for Adv.\ Studies (Sokendai), 
	Shonan Village, Japan.}
\altaffiltext{\jpl}{NASA Jet Propulsion Laboratory, California Institute of Technology, Pasadena CA, USA.}
\altaffiltext{\sinica}{Inst.\ of Astronomy and Astrophysics, Academia Sinica, Taipei, Taiwan.}
\altaffiltext{\physmath}{Kavli Institute for the Physics and Mathematics of the Universe, 
	University of Tokyo, Japan.}
\altaffiltext{\sapporo}{Department of Cosmosciences, Hokkaido University, Sapporo, Japan.}
\altaffiltext{\sendai}{Astronomical Institute, Tohoku University, Sendai, Japan}

\begin{abstract}\noindent
We present Subaru/HiCIAO $H$-band high-contrast images of the debris disk around
HIP~79977, whose presence was recently inferred from an infrared excess.  Our
images resolve the disk for the first time, allowing characterization of
its shape, size, and dust grain properties. We use angular differential imaging
(ADI) to reveal the disk geometry in unpolarized light out to a radius of
$\sim$2\arcsec, as well as polarized differential imaging (PDI) to measure the degree of scattering polarization out to $\sim$1\farcs5.  In
order to strike a favorable balance between suppression of the stellar halo and
conservation of disk flux, we explore the application of principal component
analysis (PCA) to both ADI and reference star subtraction.  This allows
accurate forward modeling of the effects of data reduction on simulated disk 
images, and thus direct comparison with the imaged disk. 
The resulting best-fit values and well-fitting intervals for the model parameters
are a surface brightness power-law slope of 
$S_\mathrm{out} = -3.2~[-3.6,-2.9]$, an 
inclination of $i = 84^\circ~[81^\circ,86^\circ]$, a high Henyey-Greenstein 
forward-scattering parameter of
$g=0.45~[0.35,0.60]$, and a non-significant disk--star offset of 
$u=3.0~[-1.5, 7.5]\,\mathrm{AU} = 
24~[-13,61]$\,mas along the line of nodes.  Furthermore, the tangential linear 
polarization along the disk rises from $\sim$10\% at 
0\farcs5 to $\sim$45\% at 1\farcs5.  These measurements paint a consistent 
picture of a disk of dust grains produced by collisional cascades and blown out
to larger radii by stellar radiation pressure.  
\end{abstract}





\section{Introduction}

Most debris disk systems are 
second-generation disks produced by the collisional destruction of 
planetesimals, although 
some of the youngest systems ($\sim$10 Myr) might also include primordial star formation material \citep[e.g.,][]{wyatt08}.  While space-based infrared (IR) surveys have identified many debris disks \citep[e.g.,][]{aum84}, though few have been spatially resolved at optical or IR wavelengths.  High-resolution images have revealed morphological disk sub-structures such as warps, geometrical offsets, and clumps \citep[e.g.,][]{kal95, hea00, kalas05, lagrange09}, possibly caused by gravitational perturbations from exoplanets. 
In particular, identifying new debris disk systems with an edge-on inclination 
is critical to efforts to study the origin and composition of their small, second-generation gaseous components \citep{rob06}.

\object{HIP 79977} (HD~146897) is a 5--10 Myr old \citep{dezeeuw99,pecaut12,song12} F2/3V member of Upper Scorpius at a distance of $123^{+18}_{-14}$ pc \citep{vanleeuwen07} which has been identified as a debris disk system with a fractional IR luminosity $L_\mathrm{IR}/L_{\star} = 5.9\times10^{-3}$ \citep{chen06, chen11}.  A simple single-temperature fit to its IR excess led \citet{chen11} to suggest a source ring radius of at least 40\,AU, with a characteristic temperature of 89\,K.  Here we present high-contrast imaging of HIP~79977 which resolves this debris disk in scattered light for the first time.
\looseness=-1

\section{Observations}

\begin{figure*}[t]
\centering
\vspace*{2mm}
\includegraphics[width=\linewidth]{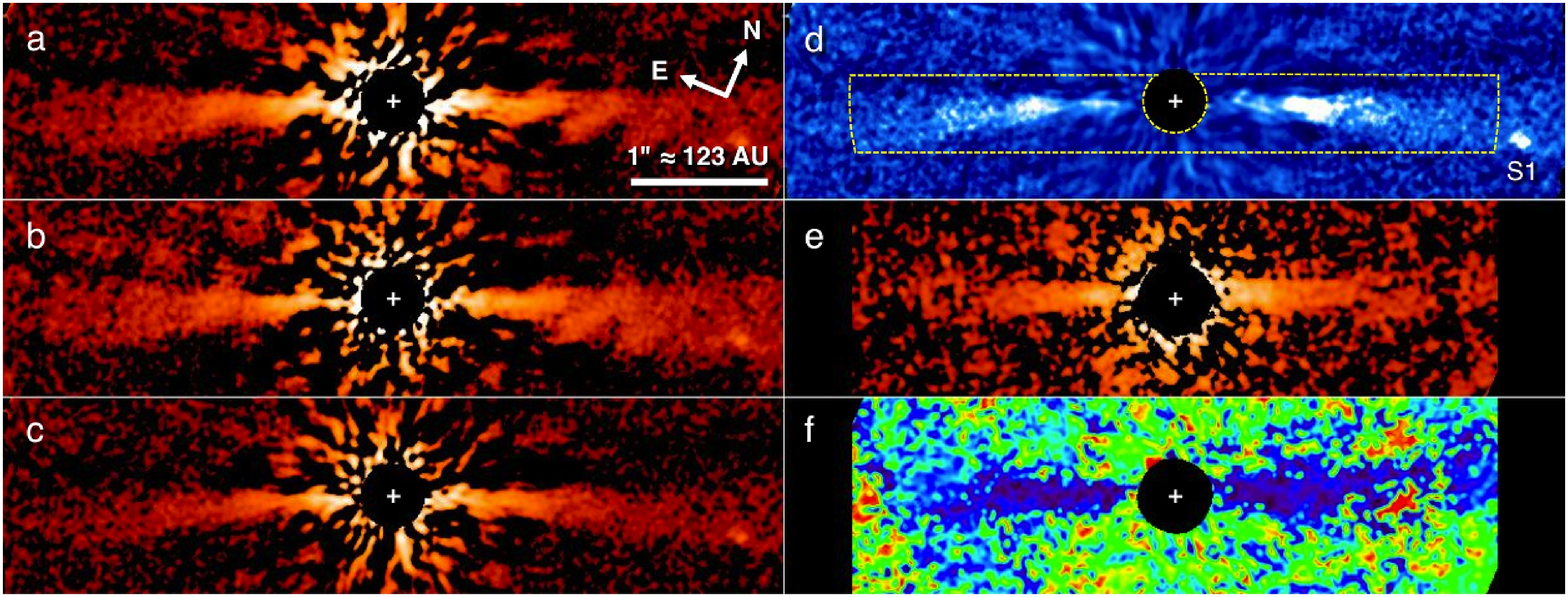}
\includegraphics[width=\linewidth]{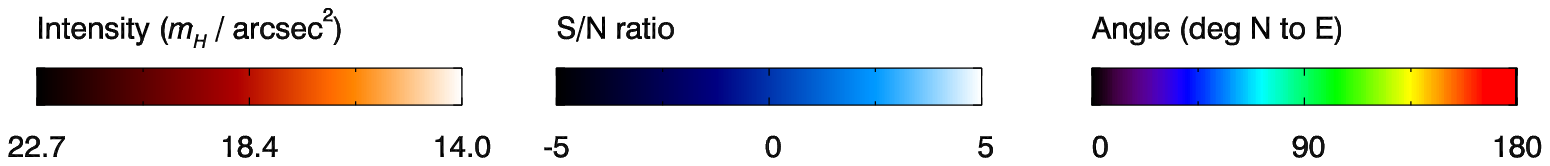}
\vspace*{1mm}
\caption{Subaru HiCIAO $H$-band images of the HIP~79977 debris disk. 
	\textbf{(a--c)} Logarithmic intensity images based on the May 2012
	data as produced by the reduction methods described in the body text.  
	\textbf{(a)}
	Classical ADI. \textbf{(b)} Conservative LOCI ($N_\delta=0.75$, $N_\mathrm{A}=
	10,000$). \textbf{(c)} PCA-assisted ADI (5 eigenmodes).
    \textbf{(d)} Signal-to-noise (S/N) map for image 
    \textbf{c} in linear stretch.  The area outlined in yellow was excluded for
    noise calculation. S1 marks a likely background star.
	\textbf{(e)} Cleaned polarized intensity image from the July 2012
    data, in logarithmic stretch. 
    \textbf{(f)} Linear polarization orientation angle. The mean orientation of 
    the disk flux
    averaged in 15-pixel boxes along the midplane is $25.7^\circ \pm 3.5^\circ$,
    consistent with the expected tangential orientation of 24.0$^\circ$.
    All images were convolved with a 5-pixel ($\approx 1$\,FWHM) diameter 
    circular aperture.
    }
\label{f:images}
\vspace*{5mm}
\end{figure*}

As part of the SEEDS survey (Strategic Exploration of Exoplanets and Disks with
Subaru/\-HiCIAO; \citealt{tamura09}), we obtained two epochs of high-contrast
imaging on \object{HIP 79977} with the HiCIAO instrument 
\citep{hodapp08} on Subaru Telescope.

The first dataset was taken in $H$-band (1.6\,$\mu$m) on 2012 May 12, 
comprising 119 frames of 
20 seconds for a total exposure time of 39.7\,min, and spanning $19.6^\circ$
of field rotation. The image rotator was operated in pupil-tracking mode to
enable angular differential imaging \citep[ADI;][]{marois06}. The plate scale was 
$9.50\pm0.02$\,mas per pixel, and the field of view 20\arcsec$\times$20\arcsec.
The AO188 adaptive optics system \citep{minowa10} provided a stable point-spread 
function (PSF) with a FWHM of 6.7~pixels $=$ 64\,mas under good weather 
conditions ($\sim$0\farcs6 DIMM seeing).  HIP~79977 saturated
out to a radius of 22~pixels~$=$~0\farcs2; no coronagraph was used.  
After correction for flatfield and 
field distortion \citep{suzuki10}, the images
were registered by fitting a Moffat profile to the PSF halo, for an
estimated registration accuracy of 0.5~pixels $=$ 5\,mas.

The second dataset was taken in two-channel polarized 
differential imaging 
(PDI) mode on 2012 July 7 under the same filter, plate scale, and saturation 
conditions, with excellent seeing (0\farcs4).
It comprises 64 frames of 30 seconds for a
total exposure time of 32\,min, and spans 17.3$^\circ$ of field rotation.  The
field of view was 10\arcsec$\times$20\arcsec.
The exposures are organized into serial batches of four
frames in which HiCIAO's half-wave plate cycled through the position angles
(0$^\circ$, 45$^\circ$, 22.5$^\circ$, 67.5$^\circ$), measuring the Stokes 
linear polarization parameters ($I+Q, I-Q, I+U, I-U$), respectively. This
allows for correction of non-common path aberrations by double-difference 
polarimetry on short time scales.  Each frame was registered by Moffat 
fitting.  While 
the intensity component of the PDI data set can also be reduced with ADI, the
resulting quality is inferior to that of the dedicated May 2012 ADI data.  We
therefore use the May 2012 data to constrain the full-intensity appearance of
the HIP~79977 disk, and the July 2012 data to obtain the matching 
polarized-intensity image.

\section{Data reduction}

\subsection{Angular differential imaging}

While the ADI technique with the LOCI algorithm 
\citep{lafreniere07} is commonly used to 
improve high-contrast sensitivity for detecting extrasolar planets 
\citep[e.g.,][]{marois10, lagrange10} and brown dwarf companions \citep[e.g.,]
[]{thalmann09}, we recently demonstrated its use for 
detecting faint circumstellar disks hidden in the
speckle halo of their host stars \citep{thalmann10,buenzli10,thalmann11}.
This ``conservative LOCI'' technique has since been widely adopted \citep{rodigas12,currie12,boccaletti12,lagrange12}. 

The benefit of powerful speckle suppression with ADI comes at the price of
partial loss of disk flux and morphological integrity \citep{milli12}. 
However, careful forward-modeling of these effects can recover the physical
disk morphology, as we demonstrated on the HR~4796~A debris disk 
\citep{thalmann11}.
Clarification is needed regarding recent discussions 
that attributed some results based on this technique to ADI artifacts rather
than to astrophysical processes
\citep{lagrange12,milli12}.  Since ADI works in concentric annuli, it cannot 
generate spurious signals at a radius where no disk flux is present.  The
presence of ``streamer''-like flux distributions therefore does imply the 
presence of physical emission at those radii, even though the streamer 
morphology itself is known to be carved from a physically smooth disk by the
ADI process.  The conclusions in \citet{thalmann11}
therefore remain valid, and we continue to use this methodology in this work.

For HIP~79977, we started with classical ADI \citep[no frame selection]{marois06}
and conservative LOCI
($N_\delta=0.75$, $N_\mathrm{A}= 10,000$).  Furthermore, we
adopted the application of principal component analysis (PCA) to ADI as a
robust and deterministic alternative to LOCI \citep{soummer12}.  
Closely following the recipe of \citet{pynpoint}, we performed PCA on the stack of 
centered, pupil-stabilized frames after subtracting the mean image and masking
the saturated central region ($r\le 22$ pixels $=$ 0\farcs2).  We then
projected each frame onto the first $n$ modes of the orthonormal set of
Karhunen-Lo\`eve eigenimages delivered by PCA, and subtracted those projections 
from the frame.
The stack of frames was then derotated and collapsed to yield the 
final image. We hereafter refer to this method as PCA-ADI.  
Through visual inspection of the resulting images, we found that $n=5$
provides an excellent trade-off between suppression of the stellar PSF halo and
conservation of disk flux.  Note that PCA-ADI with $n=0$ is identical to 
classical ADI.  To ensure linearity of our data reduction process, we use 
\texttt{mean}-based rather than \texttt{median}-based frame combination in all 
techniques \citep[cf.][]{brandt12}.  
\looseness=-1

Finally, we explored the use of PCA for PSF subtraction with reference stars. 
However, we found that for HiCIAO $H$-band data, the speckle halo varied too much
between the science and the reference targets to reach competitive contrasts. 

Scattered light from the debris disk around HIP~79977 is clearly 
detected in the May 2012 data set with all ADI-based techniques, as shown 
in Figure~\ref{f:images}a--c. The ripple
pattern visible in the first two images is a spurious residual of Subaru's 
spider diffraction pattern. Only the PCA-ADI technique proves effective at
removing this artifact and revealing the smooth slope of the debris disk.  We
therefore choose it as the benchmark technique for further analysis, as discussed
in Section~\ref{s:analysis}.  Figure~\ref{f:images}d shows the signal-to-noise
(S/N) map of the PCA-ADI image, calculated as the standard deviation in 
concentric annuli after smoothing with a 5-pixel diameter circular aperture.  
A box containing the disk flux was masked for noise calculation.

The disk presents itself almost edge-on, and thus appears as a ribbon in the
ADI images.  Most of the flux is concentrated to the South of the projected disks'
major axis, identifying this as the disk's likely near side and indicating strong 
anisotropic forward-scattering.  
The disk flux remains
statistically significant from $\sim$0\farcs3 ($\approx 40$\,AU) out to 
$\sim$2\arcsec{} ($\approx250$\,AU in projection).

\subsection{Polarized differential imaging}

We reduced the July 2012 polarized differential imaging data using the standard
procedure for SEEDS polarimetry \citep[cf.][]{tanii12}. It includes 
double-difference polarimetry using four half-wave plate position angles, as 
well as modeling and correction of instrumental polarization based on 
\citet{joos08}. The intensity PSF was fit to the $Q$ and $U$ images and subtracted
to remove contributions from a polarized stellar PSF.
The resulting polarized intensity image $P=\sqrt{(Q^2+U^2)}$ was found to contain 
spurious structures left behind by the subtraction
of imperfectly aligned PSFs, which we removed by projecting spatial derivatives of
the intensity PSF to the $Q$ and $U$ images, and subtracting them. Finally, we
noticed that RMS of the positive and negative residual speckle noise in the $Q$ 
and $U$ images added up to a positive radial halo in the $P$ image.  We
measured this noise profile at position angles that do not show disk flux, and
subtracted it in quadrature from the entire image.

The resulting $P$ image shows the debris
disk out to a separation of $r\approx1\farcs5$, where the polarized flux
level vanishes into the background (Figure~\ref{f:images}e).  The 
polarization vectors are tangentially oriented, consistent with scattering 
polarization (Figure~\ref{f:images}f). No brightness contrast between
the near and far side of the disk is evident, in accordance with the 
expectation that the excess scattered light from forward scattering on large 
grains is unpolarized.


\section{Analysis} 
\label{s:analysis}

\begin{figure}[t]
\centering
\vspace*{2mm}
\includegraphics[width=\linewidth]{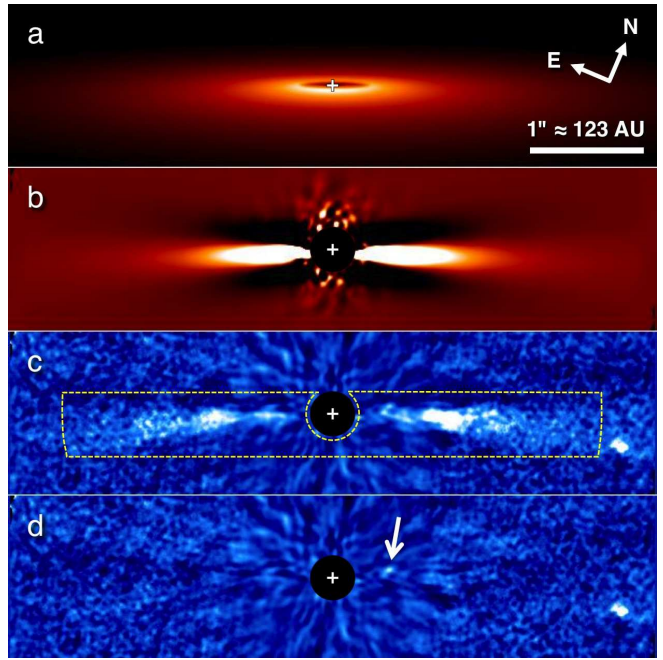}
\vspace*{1mm}
\caption{Modeling of the HIP 79977 debris disk. \textbf{(a)} Simulated 
logarithmic scattered-light image of the best-fit model disk.
\textbf{(b)} The same after applying PCA-ADI data reduction
(linear stretch to show oversubtraction). 
\textbf{(c)} Noise-normalized residual map of the PCA-ADI reduction of the May
2012 data, at a linear stretch of $[-2.5\,\sigma, 5\,\sigma]$. The evaluation
region is marked with a yellow outline.
\textbf{(d)} The same calculated after 
subtraction of the
model disk treated with PCA-ADI. The disk flux is effectively removed from the 
data (residual $\chi^2_\mathrm{min}=1095=1.10N_\mathrm{data}$). 
An unconfirmed point-like signal at 4.6\,$\sigma$ significance is
highlighted with an arrow. For visual clarity, no binning has been applied 
for panels (c,d).
}
\label{f:model}
\vspace*{2mm}
\end{figure}

\subsection{Disk modeling}
\label{s:model}

In order to extract physical disk information from our PCA-ADI image, we
generate simulated
scattered-light images of model disks using the GRaTer code (Augereau et
al. 1999). The main
free parameters of the model are the semimajor axis of the source ring $a_0$,
the inner and outer
power-law slopes $\alpha_\mathrm{in},\alpha_\mathrm{out}$ of the radial 
density distribution, the
disk--star offset $u$, the inclination
angle $i$, the position angle $\phi$, and the Henyey-Greenstein parameter $g$
characterizing the anisotropy of
the scattering behavior of the dust grains. 
For definitions we refer to
\citet{augereau99}.
Note that the measurable outer slope of the diskÕs radial surface
brightness profile is not $\alpha_\mathrm{out}$,
but $S_\mathrm{out}=\alpha_\mathrm{out}+\beta-2$, where $\beta$ represents 
the radial shape of the scale height distribution. 
While this relationship breaks down for purely edge-on disks, it holds for
the radial profile 
for our models with $i\le88^\circ$. We fix $\beta=1$ and henceforth use 
$S_\mathrm{out}$ to characterize the diskÕs outer slope. Since we do not 
resolve the gap inside the source ring, we adopt
the semimajor axis $a_0=40$\,AU ($\approx0\farcs3$) derived from 
infrared excess by \citet{chen11}.

As a first step, we determine the position angle $\phi$ of the projected 
disk's axis of symmetry. We rotate the 
PCA-ADI image from its North-up, East-left orientation by a given angle,
mirror this image about the $y$ axis, and subtract it from an unmirrored
copy, adjusting the angle to minimize the subtraction residuals.  To 
quantify the residuals, we first bin the residual image by a factor of 5 
($\approx1$ FWHM) in both dimensions, and define an evaluation region
of $N_\mathrm{data}=1025$ binned pixels framing the disk flux 
(Fig.~\ref{f:model}).  We derive a noise profile as the standard 
deviation of pixel values in concentric annuli in the residual image, 
excluding the evaluation region.  We then divide the residual image by 
this noise profile and calculate $\chi^2$ as the sum of squares of all
pixel values in the evaluation region.  

\begin{figure}[t]
\centering
\includegraphics[width=0.8\linewidth, trim=5mm 2mm 4.5mm 3mm]{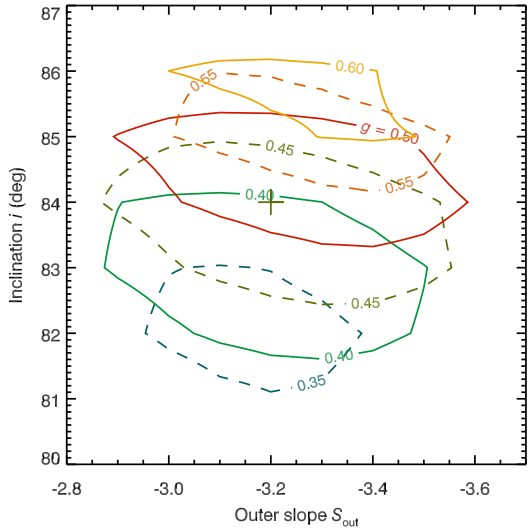}
\caption{Constraints on the disk model parameters 
$S_\mathrm{out}$ (outer slope), $i$ (inclination), and $g$ 
(Henyey-Greenstein parameter).  The contours delimit the 
parameter space of well-fitting solutions ($\chi^2 \le 
\chi^2_\mathrm{min} + \sqrt{2N_\mathrm{data}}$) for fixed values of $g$. 
The best-fit solution ($\chi^2 = \chi^2_\mathrm{min}$) lies
in the $g=0.45$ plane; it is marked with a plus sign.
}
\label{f:merit}
\vspace*{3mm}
\end{figure}

Our data reduction does not yield a residual map in which the 
pixels have fully independent normal errors, even after binning. Ê
Therefore, we do not expect the
usual $\chi^2$ thresholds to produce reliable confidence intervals, and
use a much more conservative $\chi^2$ threshold of $\sqrt{2 N_\mathrm{data}}$.
Rather than determining confidence intervals about a best-fit model,
we seek to determine the family of models consistent with our images.

For the position angles this yields best-fit value of 
$\phi=24.0^\circ$ (counter-clockwise from North) and a well-fitting range of 
$[23.7^\circ,24.3^\circ]$.  This includes HiCIAO's True North offset of 
$0.35^\circ \pm 0.02^\circ$ 
calibrated using \citet{vdMarel07}.

Next, we attempt to match the observed disk morphology by calculating a 
grid of model disks with $i=[77^\circ\ldots88^\circ]$, 
$S_\mathrm{out}=[-2.8\ldots-3.7]$, and 
$g=[0.00\ldots0.80]$.  Rather than implanting the models into 
the raw data as we did in \citet{thalmann11},
we exploit the deterministic nature of PCA-ADI to forward-model the exact
effect of our data reduction process on the model disk 
\citep[cf.][]{soummer12}. We subtract a scaled version of 
this processed disk image from the data, choosing the scale factor to 
minimize residual $\chi^2$ as defined above.  This yields best-fit model
parameters and well-fitting parameter ranges of  $i=84^\circ$ $[81^\circ,86^\circ]$, 
$S_\mathrm{out}=-3.2$ $[-3.6,-2.9]$, $g=0.45$ $[0.35, 0.60]$, and a ratio of 
total disk flux to stellar flux of $3.3 [2.9,3.8] \times10^{-3}$.  
The minimum $\chi^2$ achieved is $1142=1.1N_\mathrm{data}$.
The well-fitting family is characterized graphically in 
Fig.~\ref{f:merit} and summarized in Table~\ref{t:results}.  
The parameters $i$ and $g$ are somewhat degenerate; higher inclinations mimic 
enhanced forward scattering.

In a final step, we explore small non-zero eccentricities of the disk 
while keeping the best-fit values of $i$, $S_\mathrm{out}$, and 
$g$ fixed.  We approximate eccentric model disks by translating a
circularly symmetric disk by an offset $u$ along the line of 
nodes (roughly corresponding to the projected disk's major 
axis), for an eccentricity of $e = u/a$.
Positive values of $u$ are assigned to offsets to the West.  
We find a best-fit offset of $u=+2.4\,\mathrm{AU} = 
20$\,mas, with a well-fitting interval of $[-1.5,+6.3]$\,AU.
While
the best-fit offset corresponds to an eccentricity of $e=0.06$, the 
well-fitting range is consistent with zero eccentricity.

Figure~\ref{f:model} illustrates this best-fit model and its 
subtraction residual. The model fully explains the observed image
morphology down to the noise threshold.

\begin{table}[t]
\vspace*{-3mm}
\centering
\caption{Summary of Properties and Results for HIP~79977}
\label{t:results}
\vspace*{-1mm}
\begin{tabular*}{\linewidth}{@{}l@{}r@{}r}   
\\
\hline
\hline
Fixed disk model parameters	& adopted	& constraints\\
\hline
Source ring semimajor axis $a_0$ (AU, \arcsec)\noteone	&\hspace*{-3mm}$40\approx0\farcs33$ 
		& $<60\approx0\farcs49$\\
Inner brightness distribution slope $\alpha_\mathrm{in}$\qquad\qquad\qquad &$20$\\
Scale height radial shape parameter $\beta$	& $1$%
\vspace{1mm}\\
Optimized disk model parameters & best fit	& well-fitting\\
\hline
Minor axis position angle $\phi$ ($^\circ$)	& $24.0$		& $[23.7,24.3]$\\
Inclination $i$ ($^\circ$)		& $84$			& $[81, 86]$\\
Henyey-Greenstein parameter $g$			& $0.45$		& $[0.35, 0.60]$\\
Disk--star offset $u$ (AU)\notetwo{}			& $3.0$ & $[-1.5, +7.5]$\\
Eccentricity $e=\mathrm{abs}(u/a_0)$			& 0.06	& $[0, 0.16]$\\
\multicolumn{3}{@{}l@{}}{Outer brightness distribution slope 
	$S_\mathrm{out} = \alpha_\mathrm{out} + \beta -2$:} \\
--- model disk, true slope		& $-3.2$ &	$[-3.6, -2.9]$\\
--- model disk, after PCA-ADI	& $-2.6$ 	& $[-2.9, -2.3]$\\
--- measured, after PCA-ADI				& $-2.7$		& $[-2.8, -2.6]$\\
Disk/star flux contrast, $H$-band	& $0.0033$		& \hspace*{-7mm}$[0.0029, 0.0038]$%
\vspace{1mm}\\
Degree of scattering polarization	& \hspace*{-3mm}measured & 1\,$\sigma$ interval\\
\hline
--- at $0\farcs5 \approx 62$\,AU (\%)		&$10$ 	& $[5, 20]$\\
--- at $1\farcs0 \approx 123$\,AU (\%)		&$35$ 	& $[20, 45]$\\
--- at $1\farcs5 \approx 185$\,AU (\%)		&$45$ 	& $[30, 60]$%
\vspace{3mm}\\
\multicolumn{3}{@{}p{\linewidth}}{%
	\textsc{Notes.} The given ranges for parameters $i$, $g$, 
	$S_\mathrm{out}$, $e$, and $u$ represent the minimum and maximum values
	of the well-fitting family of disk models defined by $\chi^2 \le 
	\chi^2_\mathrm{min} + \sqrt{2N_\mathrm{data}}$; 
	see text and Figure~\ref{f:model} for details.
	(1) S/N ratios near the 
	inner working angle are too
	low for optimization of $r_0$; we adopt the value of 40\,AU from 
	\cite{chen11}.
	(2) The eccentric disk is approximated as a symmetric disk offset from
	the star along the line of nodes.
	Positive values of $u$ indicate a disk center to the West of the star. 
	}
\end{tabular*}
\vspace*{4mm}
\end{table}

\subsection{Surface brightness profiles}

\begin{figure}[t]
\centering
\includegraphics[width=\linewidth, trim=4mm 0mm 0mm 0mm]{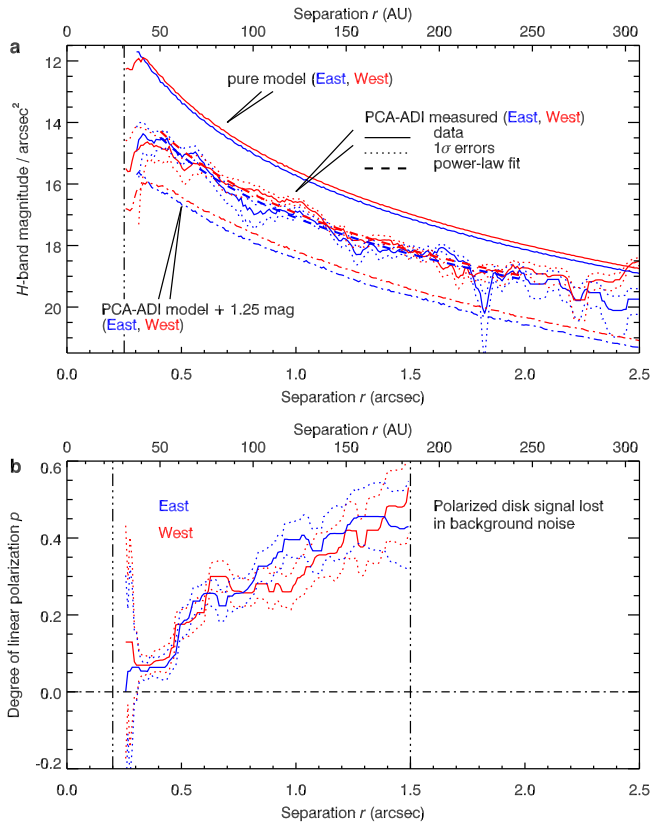}
\vspace*{1mm}
\caption{Intensity and polarization profiles of the HIP~79977
debris disk.  
\textbf{(a)} Surface brightness profiles measured in wedges of $9^\circ$
along the eastern and western traces of the debris disk image between
position 
angles $[113^\circ$, $122^\circ]$ and $[286^\circ$, $295^\circ]$, 
respectively. 
The image was smoothed with a 5-pixel diameter circular aperture prior
to extraction. The solid curves with dotted error sheaths represent the
profiles measured on the May 2012 data processed with PCA-ADI. The errors
were obtained from evaluating a number of $9^\circ$ wedges placed in the
background at disk-free position angles. Power-law fits to those profiles
are overplotted as dashed curves. The fitted slopes are $-2.6\pm0.1$ on both 
sides of the disk. Furthermore, the surface brightness profiles of the 
best-fit model disk before (top solid curves) and after PCA-ADI treatment 
(bottom dash-dotted curves) are shown for comparison. The latter are
offset downwards by $+1.25$\,mag/arcsec$^2$ to avoid blending with the 
profiles measured from the data. The pure model profiles
are not artificially offset; the difference of $1$--$2.5$\,mag/arcsec$^2$
with respect to the measured disk profile represents the flux loss 
incurred in PCA-ADI. \textbf{(b)} Degree of scattering polarization $p$ at a 
function of separation, calculated from the polarized intensity image ($P$)
measured in the July 2012 data and the full intensity $I$ image of the
best-fit disk model derived from the May 2012 data.  The profiles are 
measured in $6^\circ$ wedges and smoothed
with a median filter of 11 pixels ($\approx 2$\,FWHM). The polarized disk
flux is lost in the background noise beyond  1\farcs5 
(cf.\ Figure~\ref{f:images}e).
}
\label{f:profpol}
\vspace*{5mm}
\end{figure}

As a complementary approach, we directly measure the surface brightness
profile of the debris disk.  To this end, we convolve the PCA-ADI image
with a 5-pixel diameter circular aperture and average azimuthally
in radial sectors of 9$^\circ$ centered on the position angles of the
traces of the imaged disk. The results are shown in Figure~\ref{f:profpol}a.
The profiles are well-described by a power law with a slope of 
$S_\mathrm{data} = -2.7 \pm 0.1$.  

While the best-fit model disk found in Section~\ref{s:model} has an intrinsic
surface brightness slope of $S_\mathrm{out}=-3.2$, the forward-modeling
of the effects of PCA-ADI data reduction on that disk yields a softened 
slope of $S=-2.6\pm0.3$, consistent with our data.

\subsection{Linear polarization}

We measure the polarized intensity profile in sectors of $6^\circ$
centered on the disk traces in the fully reduced $\mathit{P}$ image, from
which the RMS halo from residual speckle noise has been subtracted.  We 
confirm that the linear polarization is oriented tangentially with respect
to the star (Fig.~\ref{f:images}f), as is expected for scattering 
polarization. To calculate the 
degree of scattering polarization, we divide the polarized intensity profile
by the full intensity profile of the best-fit model disk evaluated in the 
same sectors. 

The results are presented in Figure~\ref{f:profpol}b.  We find that the
scattering polarization increases from $\sim$10\% at 0\farcs5 to $\sim$45\%
at 1\farcs5. This might be due to 
scattering angles at large separations for the considered position angles
being constrained to $\sim$90$^\circ$, where scattering produces the 
highest amount of polarization.  Close to the star, anisotropic 
forward-scattering at angles $>90^\circ$ yields a surplus of unpolarized
light, diluting the degree of polarization.

\subsection{Point sources}

Two faint point sources are clearly visible within 3\arcsec{} of HIP 79977
in the ADI data, one of which is shown in Fig.~\ref{f:images}d. 
A preliminary astrometric
analysis based on marginal detections in archival Gemini NICI data
identifies both as likely unrelated background stars, though more
precise astrometry is needed for a decisive result.  Furthermore, 
subtraction of the best-fit model disk leaves behind a
point-like signal of $4.6\sigma$ significance at a separation of 
0\farcs5 (cf.\ Fig.~\ref{f:model}d).  If confirmed, such a signal may
represent scattered light from a localized clump in the debris disk, 
or the thermal emission from a young planet of 3--5 Jupiter masses
(based on \citealt{baraffe03} and partial self-subtraction correction).
While we detect no warps or gaps in the disk, this 
does not preclude the presence of a planet of the proposed mass and 
separation range orbiting outside the disk's source ring \citep{thebault12}.  


\section{Discussion}

The results of our imaging characterization of the debris disk around 
HIP~79977 fit very well with the emerging standard picture of debris
disks, where dust is produced in a ring of colliding planetesimals and 
then distributed inward and outward by radiation forces.  The slope of 
the scattered-light surface brightness is measured to be 
within $[-3.6,-2.9]$, which is consistent with the theoretical 
value of $-3.5$ typical for a radiation pressure driven eccentricity 
distribution of particles near blow-out size \citep{krivov06, strubbe06,
thebault07}.  No significant eccentricity of the disk as a whole is 
measured, potentially setting HIP~79977 apart from other disks like
those around Fomalhaut \citep{chiang09, acke12} and HR~4796~A 
\citep{thalmann11}.
However, future observations of HIP~79977 resolving the source ring
may reveal asymmetries below the current uncertainty threshold of 
$e\le0.16$.  

The derived Henyey-Greenstein parameter of $g\approx0.45$ is consistent 
with measurements of cometary dust \citep{kolokolova04}.  Likewise, a
maximum degree of polarization of $45 \pm 15\%$ in $H$-band agrees well
with an extrapolation to higher phase angles of observations
of high-polarization comets like comet Hale-Bopp \citep{hasegawa97}.

As a caveat, we note that other well-fitting disk architectures may 
exist outside the limitations of our model assumptions, and that the
possible degeneracies of PCA-ADI images of disks have not been
investigated in depth.


\acknowledgements
We thank 
Jean-Charles Augereau for his
GRaTer code, and the anonymous referee for helpful comments.
J.C.\ is 
supported by the U.S.\ National Science Foundation under Award 
No.~1009203.
The authors wish to recognize and acknowledge the significant 
cultural role and reverence that the summit of Mauna Kea has always had 
within the indigenous Hawaiian community.  We are most fortunate to have
the opportunity to conduct observations from this mountain.

{\it Facilities:} \facility{Subaru (HiCIAO, AO188)}.

\clearpage

\end{document}